%% file: paper.tex
\documentclass[sigconf]{acmart}

\pdfoutput=1

\makeatletter                   
\def\mdseries@tt{m}             
\makeatother                    

\usepackage[plain]{fancyref}
\usepackage[draft=true]{minted} 
\usepackage{color}
\usepackage{cleveref}
\usepackage{breakurl}           
\usepackage{graphicx}

\usepackage{subfigure}

\crefformat{section}{\S#2#1#3}
\crefname{figure}{figure}{figures}
\crefname{table}{table}{tables}

\renewcommand\footnotetextcopyrightpermission[1]{} 
\setcopyright{none}

\begin{document}

\title{
Characterizing Co-located Datacenter Workloads:\\
An Alibaba Case Study
}

\author{
    {\rm Yue Cheng, Zheng Chai, Ali Anwar$^{\dagger}$}\\
    {{\it George Mason University $^{\dagger}${IBM Research--Almaden}}}
}

\begin{abstract}

Warehouse-scale cloud datacenters co-locate workloads with different
and often complementary characteristics for improved resource
utilization. To better understand the challenges in managing such
intricate, heterogeneous workloads while providing quality-assured
resource orchestration and user experience, we analyze
Alibaba's co-located workload trace, the first publicly available
dataset with precise information about the category of each job. Two
types of workload---long-running, user-facing, containerized
production jobs, 
and transient, highly dynamic, non-containerized, and non-production
batch jobs---are running on a shared cluster of 1313 machines. Our
multifaceted analysis reveals insights that we believe are useful for
system designers and IT practitioners working on cluster management
systems.

\end{abstract}

\maketitle

\thispagestyle{empty}

\input{intro}

\input{background}
\input{overall}
\input{container}
\input{batch}

\input{colocation}

\input{conclusion}


\label{startofrefs} 

{
\bibliographystyle{acm}
\bibliography{alibaba}
}

%

\end{document}

%% file: intro.tex
\pdfoutput=1

\section{Introduction}


While modern datacenter management systems play a central role in
delivering quality-assured cloud computing services, warehouse-scale
datacenter infrastructure often comes with a tremendously huge cost
of low resource utilization. Google's production cluster trace
analysis reports that the overall utilization are between
20--40\%~\cite{google_dc} most of the time.
Another study~\cite{ec2_dasc11} observes that a fraction of Amazon
servers hosting EC2 virtual machines (VMs) show an average CPU
utilization of $7\%$ over  one week. 

To improve resource utilization and thereby reduce costs, leading
cloud infrastructure operators such as Google and Alibaba
\emph{co-locate}
transient batch jobs with long-running, latency-sensitive,
user-facing jobs~\cite{container_mgmt_acmqueue16, borg_eurosys15} on
the same cluster.
%
Workload co-location resembles hypervisor-based server
consolidation~\cite{xen_sosp03} but at massive datacenter scale. At
its core, the driving force is what is called a Datacenter Operating
System~\cite{dcos_hotcloud11}, managing job scheduling, resource
allocation, and so on.
As one example, Google's Borg~\cite{borg_eurosys15}
adopts the workload co-location technique by leveraging resource
isolation provided by Linux containers~\cite{lxc}.  

Workload co-location has become a common
practice~\cite{container_mgmt_acmqueue16, borg_eurosys15} and there
have been studies focusing on enabling more efficient co-location
based cluster management~\cite{paragon_asplos13, quasar_asplos14,
slo_socc17, bubbleup_micro11, omega_eurosys13, cpi2_eurosys13}.
To better facilitate the understanding of interactions among the
co-located workloads and their real-world operational demands
Alibaba recently released a cluster usage and co-located workload
dataset~\cite{alitrace}, which was collected from Alibaba's
production cluster in a 24-hour period.

We perform a characterization case study targeting Alibaba's
co-located long-running and batch job workloads 
across several dimensions. We analyze the resource request and
reservation patterns, resource usage, workload dynamicity, straggler
issues, interaction and interference of co-located workloads, among
other aspects. 
While confirming old issues still \emph{surprisingly} persist (e.g.,
the straggler issues for batch workloads) that have long been
observed in other work~\cite{dolly_nsdi13, late_osdi08}, we make
several unique insightful findings. Some of them may be specific to
the Alibaba infrastructure, but we believe the generality is critical
and applicable to designers, administrators, and users of co-located
resource management systems.
Our key findings are summarized as follows:

\textbf{Overbooking, over-provisioning, and over-commitment.}
Overbooking happens at long-running container deployment phase but
just sparsely, only for few jobs that may not have strict overbooking
requirements (e.g., CPU core sharing).
Over-provisioning mainly happens for long-running containerized jobs
for accommodating potential load spikes; but most time long-running
jobs are CPU-inactive, leaving co-located batch jobs ample
opportunity space for elastic resource over-commitment for improved
cluster resource utilization. Long-running jobs are more
memory-demanding, hence yielding an overall cluster memory
utilization higher than that of CPU. 

\textbf{Notoriously persistent old issues.}
Old issues such as poorly predicted resource usage and stragglers for
batch job workloads still persist at Alibaba's datacenters. Accurate
resource usage estimation for batch jobs is not an urgently demanding
mission since non-production batch jobs can always over-commit
resources that are reserved but not utilized by long-running
production jobs.  On the other hand, straggler issues still exist and
demand fixings at either administrator or developer side.

\textbf{Co-location implications.}
High resource sharing means intricate resource contentions at
different levels of the software stack, and potentially high
performance interference.
Our analysis also reveals evidences implying that Alibaba's
workload-specific schedulers for long-running and batch jobs may not
be as cohesively coordinated as they should, stressing the need for a
more integrated, co-location-optimized solution.



%% file: background.tex
\pdfoutput=1

\vspace{-8pt}
\section{Background and Related Work}
\label{sec:trace}
\vspace{-5pt}

\textbf{Cluster trace studies.}
In 2011, Google open-sourced the first publicly available cluster
trace data~\cite{google_trace} spanning several clusters.
Reiss et~al.~\cite{google_trace_socc12} study the heterogeneity and
dynamicity properties of the Google workloads. Other
works~\cite{google_cluster12, google_icac12, google_icppw12} focus
their studies on different aspects of the Google trace.
Alibaba, the largest cloud service provider in China, released
their cluster trace~\cite{alitrace} in late 2017.  Different from the
Google trace, the Alibaba trace contains information about the two
co-located container and batch job workloads, facilitating better
understanding of their interactions and interferences.
Lu~et~al.~\cite{alitrace_bigdata17} perform characterization of the
Alibaba trace to reveal basic workload statistics.  Our study is
focused on providing a unique and microscopic view about how the
co-located workloads interact and impact each other. 

\begin{figure}[t]
\begin{center}
\includegraphics[width=0.4\textwidth]{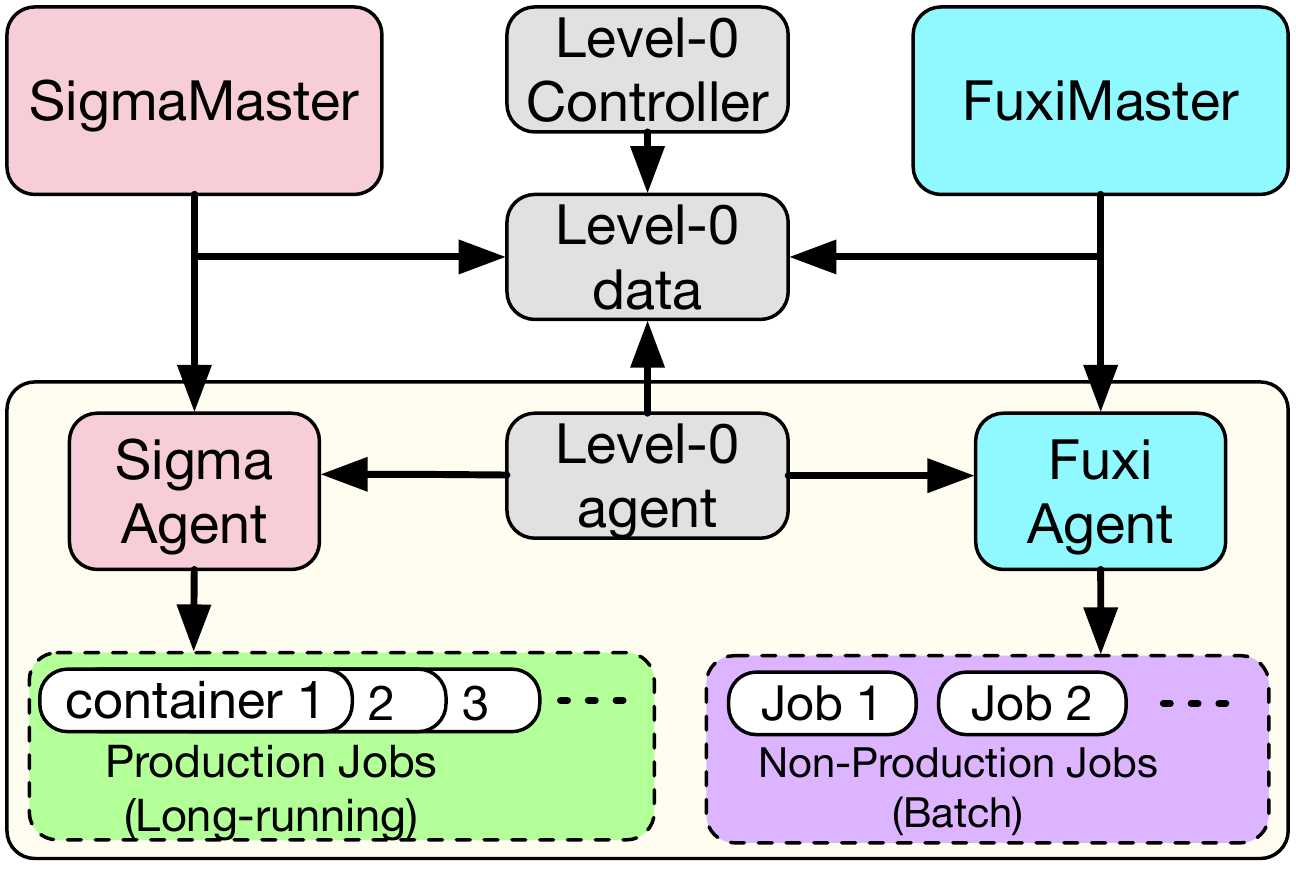}
\vspace{-7pt}
\caption{{\small
Alibaba cluster management system architecture. Each server machine has
an instance of Sigma agent and Fuxi agent which coordinate with the
Level-0 management component to enable Sigma and Fuxi schedulers to
work together.
}}
\vspace{-15pt}
\label{fig:arch}
\end{center}
\end{figure}

\textbf{Cluster management systems.}
A series of state-of-the-art cluster management systems (CMSs)
support co-located long-running and batch services. Monolithic CMSs
such as Borg~\cite{borg_eurosys15}\footnote{
Technically, Borg partitions the whole cluster into cells and assigns
a replicated scheduler for each cell that provides a certain set of
services, e.g., batch services.
} (and its open-sourced implementation Kubernetes~\cite{kubernetes})
and Quasar~\cite{quasar_asplos14} use a centralized resource
scheduler for performing resource allocation and management.
Two-level CMSs such as Mesos~\cite{mesos_nsdi11} adopts a
hierarchical structure where a Level-0 resource manager is used to
jointly coordinate multiple Level-1 application-specific schedulers.
Other CMSs achieve low-latency scheduling by using approaches such as
shared-state parallel schedulers~\cite{omega_eurosys13}.

\textbf{Alibaba's CMS and co-located workloads.}
Alibaba's CMS resembles the architecture of a two-level CMS,
as shown in Figure~\ref{fig:arch}.
The logical architecture consists of three component:
(1)~a global Level-0 controller. 
(2)~a Sigma~\cite{alibaba_sigma} scheduler that manages the
long-running workload, 
and (3)~a Fuxi~\cite{fuxi_vldb14} scheduler that manages the batch
workloads. 
Sigma is responsible for scheduling \textbf{online service
containers~\cite{pouch} for the production jobs}. 
These online services are user-facing, and typically require low
latency and high performance.  
Sigma has been used for large-scale container deployment purposes by
Alibaba for several years. It has also been used during the Alibaba
Double 11 Global Shopping Festival. We analyze the long-running
workloads in Section~\ref{sec:container}.
Fuxi, on the other hand, manages \textbf{non-containerized
non-production batch jobs}.  Fuxi is used for vast amounts of data
processing and complex large-scale computing type applications. Fuxi
employs data-driven multi-level pipelined parallel computing
framework, which is compatible with
MapReduce~\cite{mapreduce_osdi04},
Map-Reduce-Merge~\cite{mrm_sigmod07}, and other batch programming
modes. We study the batch workloads in Section~\ref{sec:batch}.
The Level-0 mechanism coordinates and manages both types of workloads
(Sigma and Fuxi) for coordinated global operations. Particularly the
Level-0 controller performs four important functionalities. (1)
manages colocation clusters, (2) performs resource matching between
each scheduling tenant, (3) strategies for everyday use and use
during large-scale promotions, and (4) performs exception detection
and processing. Different types of workloads were
running on separate clusters before 2015, since when Alibaba has been
making effort to co-locate them on shared clusters.
In Section~\ref{sec:col} we analyze Alibaba's workload co-location and its
implication.


\textbf{Alibaba cluster trace.}
The Alibaba cluster trace captures detailed statistics for the
co-located workloads of long-running and batch jobs over a course
of 24 hours. The trace consists of three parts: 
(1)~statistics of the studied homogeneous cluster of 1313~machines,
including each machine's hardware configuration\footnote{
Over $99.6\%$ machines have the same hardware composition (64-core CPU,
normalized memory capacity $0.69$, and normalized disk capacity $1$,
except a few with slightly different memory capacity).
}, and the runtime $\{$CPU, Memory, Disk$\}$ resource usage for
a duration of 12 hours (the $2^{nd}$ half of the 24-hour period);
(2)~long-running job workloads, including a trace of all container
deployment requests and actions, and a resource usage trace for 12
hours;
(3)~co-located batch job\footnote{
Each job is a directed acyclic graph (DAG), having one or more tasks;
each task has multiple instances; all instances within a task execute
the same binary; instance is the smallest scheduling unit of a batch
job scheduler.
} workloads, including a trace of all batch job requests and actions,
and a trace of per-instance resource usage over 24 hours.  Unlike the
Google trace~\cite{google_trace_socc12} that lacks precise
information about exact purpose of individual jobs, the Alibaba trace
well compensates this by tracing the two different workloads
separately, thus offering researchers visibility of real-world
operational demands of co-located workloads.

%% file: overall.tex
\pdfoutput=1

\section{Overall Cluster Usage}
\label{sec:overall}

\begin{figure}[t]
\begin{center}
\subfigure[CPU usage.] {
\includegraphics[width=.235\textwidth]{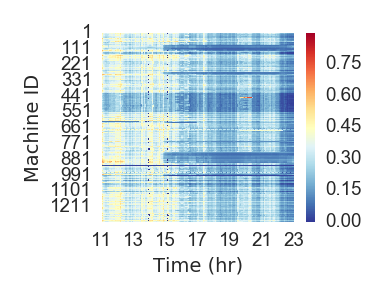}
\label{fig:overall_cpu}
}
\hspace{-1.5em}
\subfigure[Memory usage.] {
\includegraphics[width=.235\textwidth]{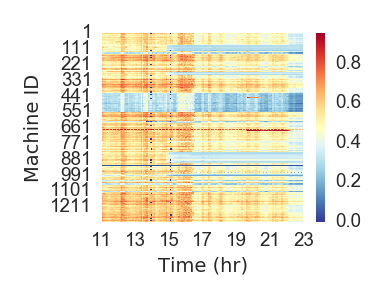}
\label{fig:overall_mem}
}
\vspace{-15pt}
\caption{{\small
Cluster resource usage (hour~11--23).
}}
\label{fig:overall}
\end{center}
\vspace{-15pt}
\end{figure}

\begin{figure}[t]
\begin{center}
\subfigure[Average CPU usage.] {
\includegraphics[width=.235\textwidth]{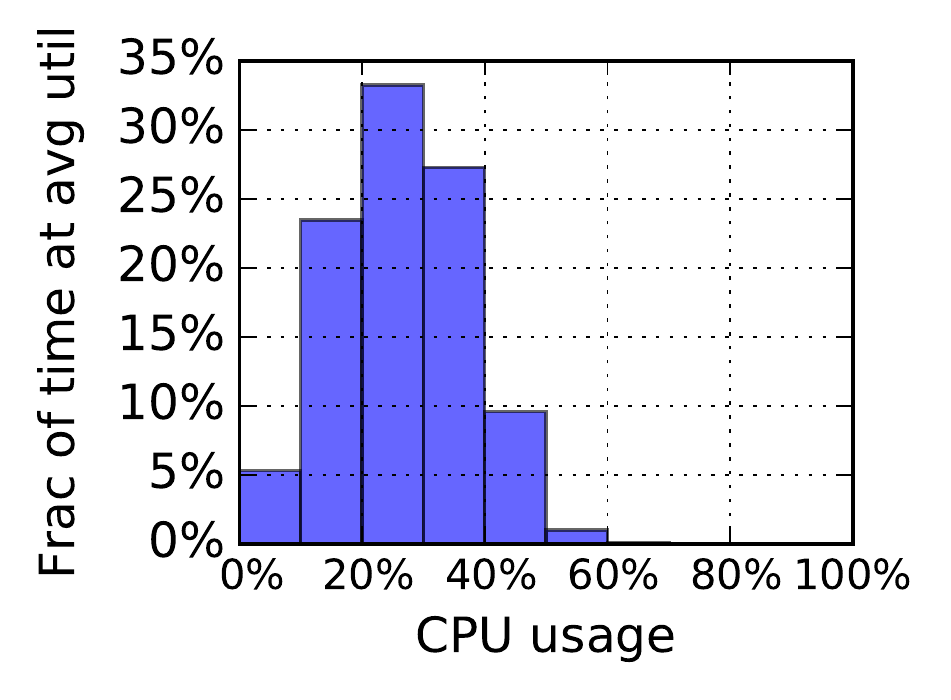}
\label{fig:cpu_hist}
}
\hspace{-1.2em}
\subfigure[Average memory usage.] {
\includegraphics[width=.235\textwidth]{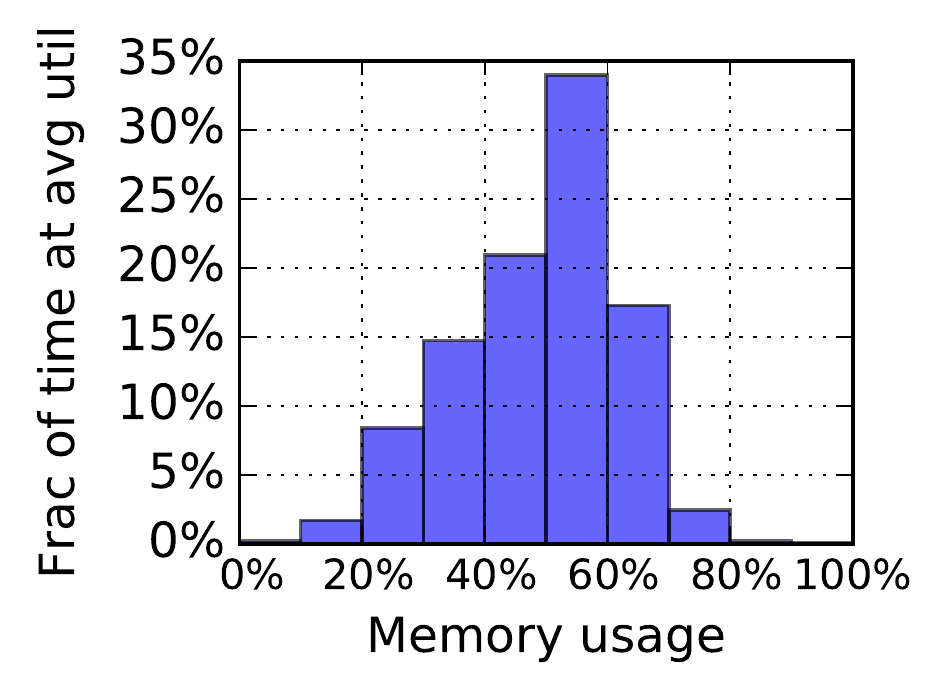}
\label{fig:mem_hist}
}
\vspace{-15pt}
\caption{{\small
Histogram of average resource usage.
}}
\label{fig:hist}
\end{center}
\vspace{-15pt}
\end{figure}

We start with the overall cluster usage to understand the aggregated
workload characteristics. Figure~\ref{fig:overall} shows the
per-machine resource usage temporal pattern across the 12-hour
duration. Cluster CPU utilization (Figure~\ref{fig:overall_cpu}) is
at medium level ($40\%$--$50\%$) for the first 4 hours but decreases
during the rest time, while memory utilization
(Figure~\ref{fig:overall_mem}) is above $50\%$ in majority of the
time. This can also be reflected from Figure~\ref{fig:hist}, which
shows the average usage distribution. The cluster spends over $80\%$
of its time running between $10\%$--$30\%$ CPU usage
(Figure~\ref{fig:cpu_hist}).  Average memory usage
(Figure~\ref{fig:mem_hist}) is relatively higher and in over $55\%$
of the time the machines have a memory usage above $50\%$. 
Google trace analysis~\cite{google_trace_socc12} reports
that the CPU and memory usage were capped at $60\%$ and $50\%$,
respectively. In contrast, at Alibaba, memory tends to be more
precious with over half of the capacity consumed for over half of the
time.

%% file: container.tex
\pdfoutput=1



\section{Long-Running Job Workloads}
\label{sec:container}

\begin{figure}[t]
\begin{center}
\includegraphics[width=0.45\textwidth]{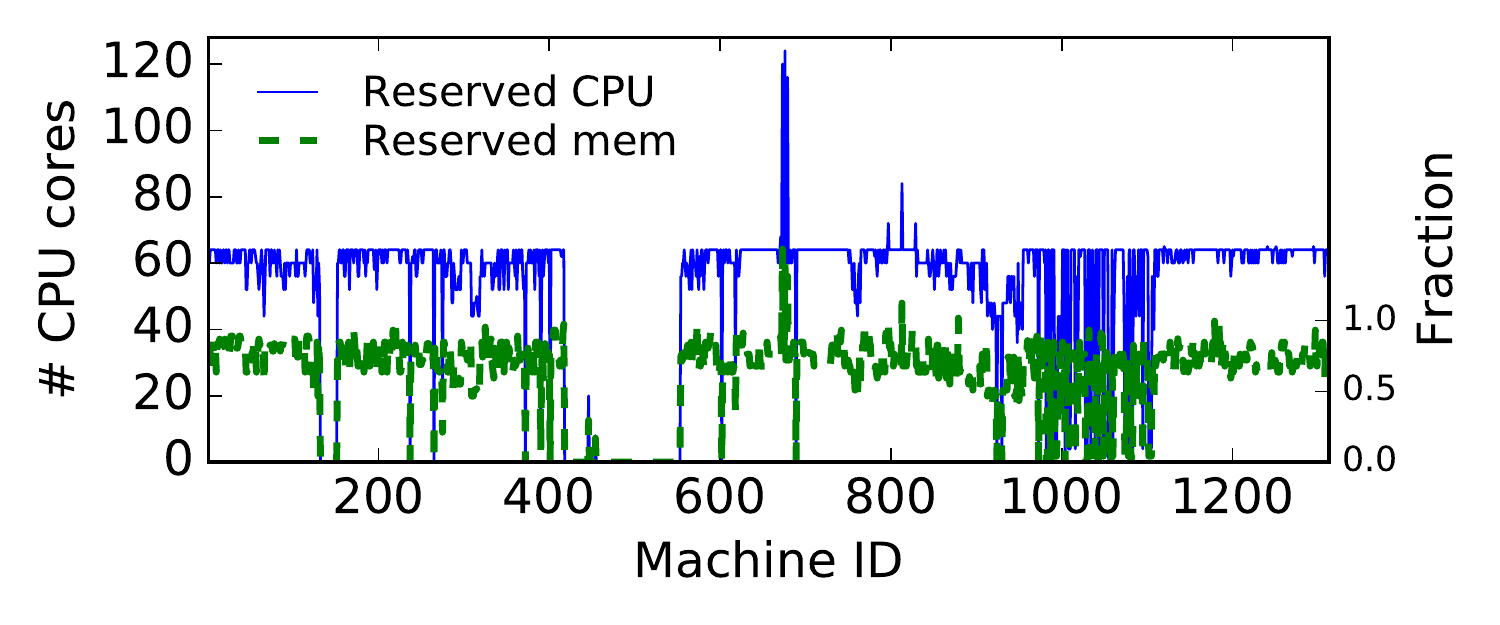}
\vspace{-15pt}
\caption{{\small
Distribution of reserved resources at container creation time across
the cluster (note reserved memory capacity (normalized) are shown on
secondary Y-axis).
}}
\vspace{-15pt}
\label{fig:c_alloc}
\end{center}
\end{figure}

\begin{figure}[t]
\begin{center}
\includegraphics[width=0.45\textwidth]{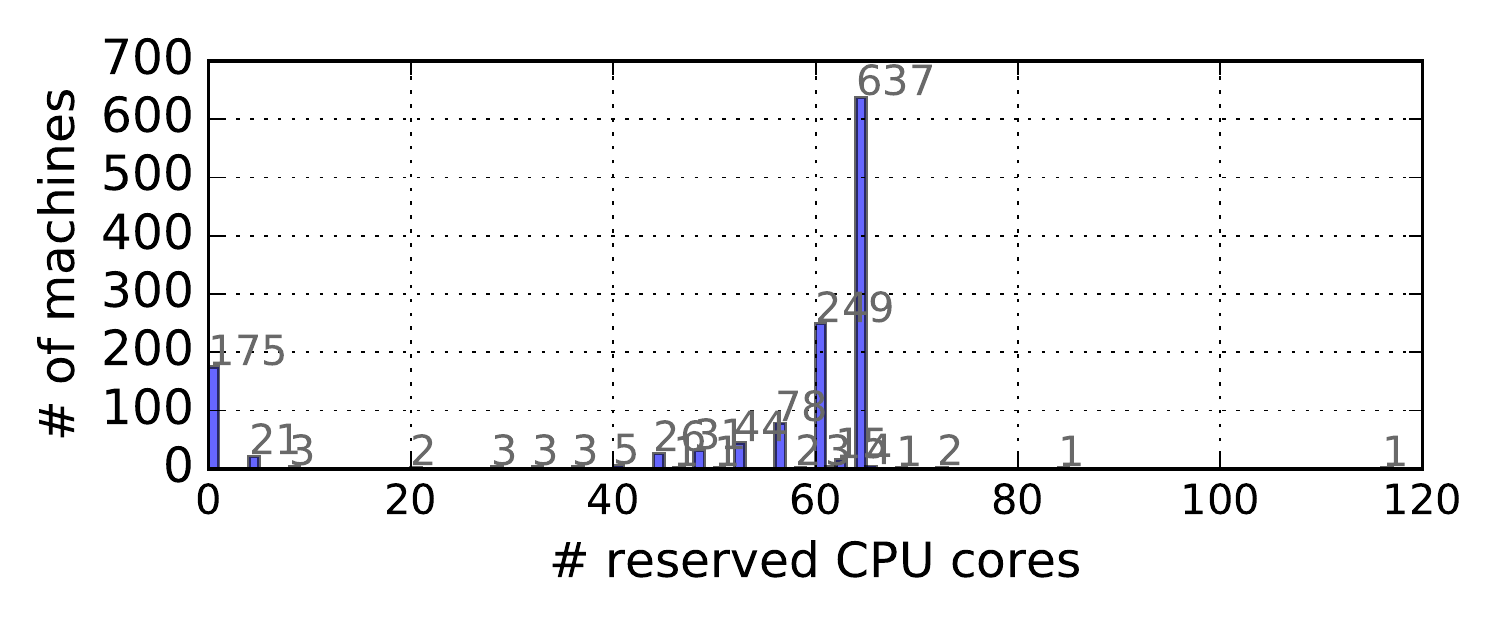}
\vspace{-15pt}
\caption{{\small
Histogram of reserved CPU cores.
}}
\vspace{-15pt}
\label{fig:lr_cpu_req_dist}
\end{center}
\end{figure}

\begin{figure}[t]
\begin{center}
\includegraphics[width=0.45\textwidth]{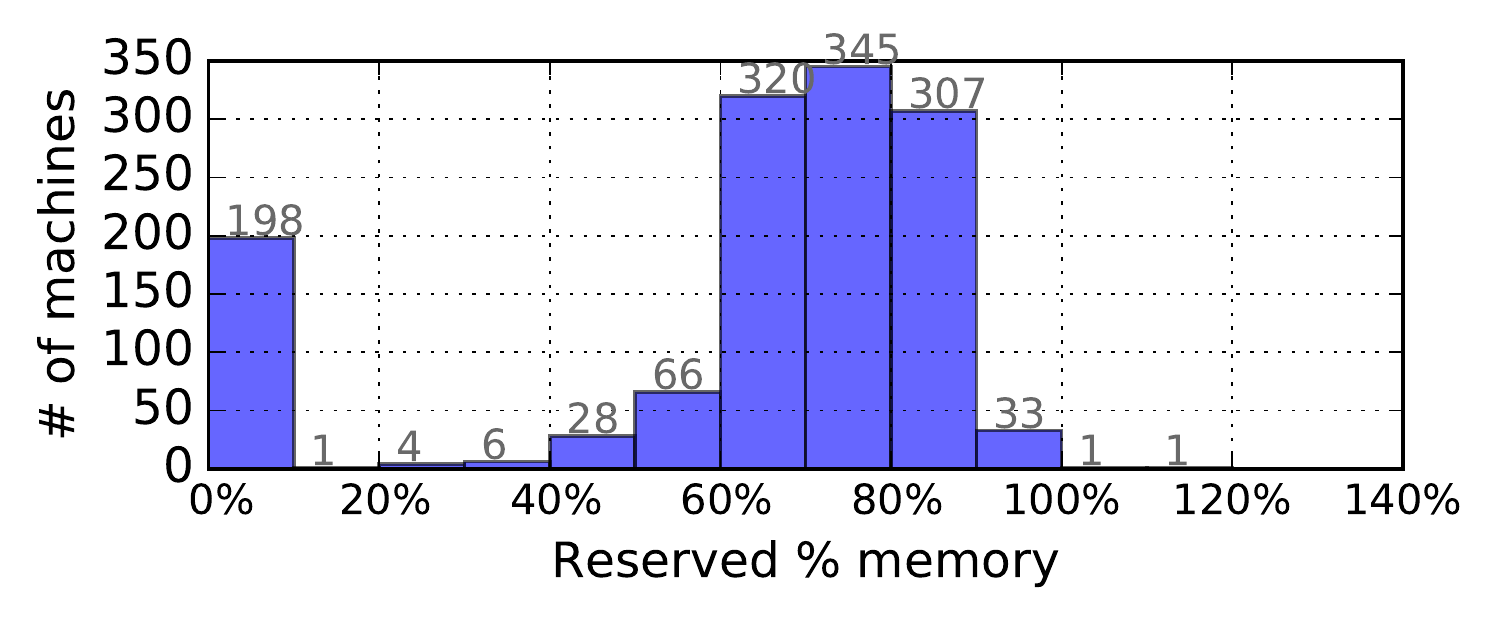}
\vspace{-15pt}
\caption{{\small
Histogram of reserved memory.
}}
\vspace{-15pt}
\label{fig:lr_mem_req_dist}
\end{center}
\end{figure}

\begin{figure}[t]
\begin{center}
\includegraphics[width=0.42\textwidth]{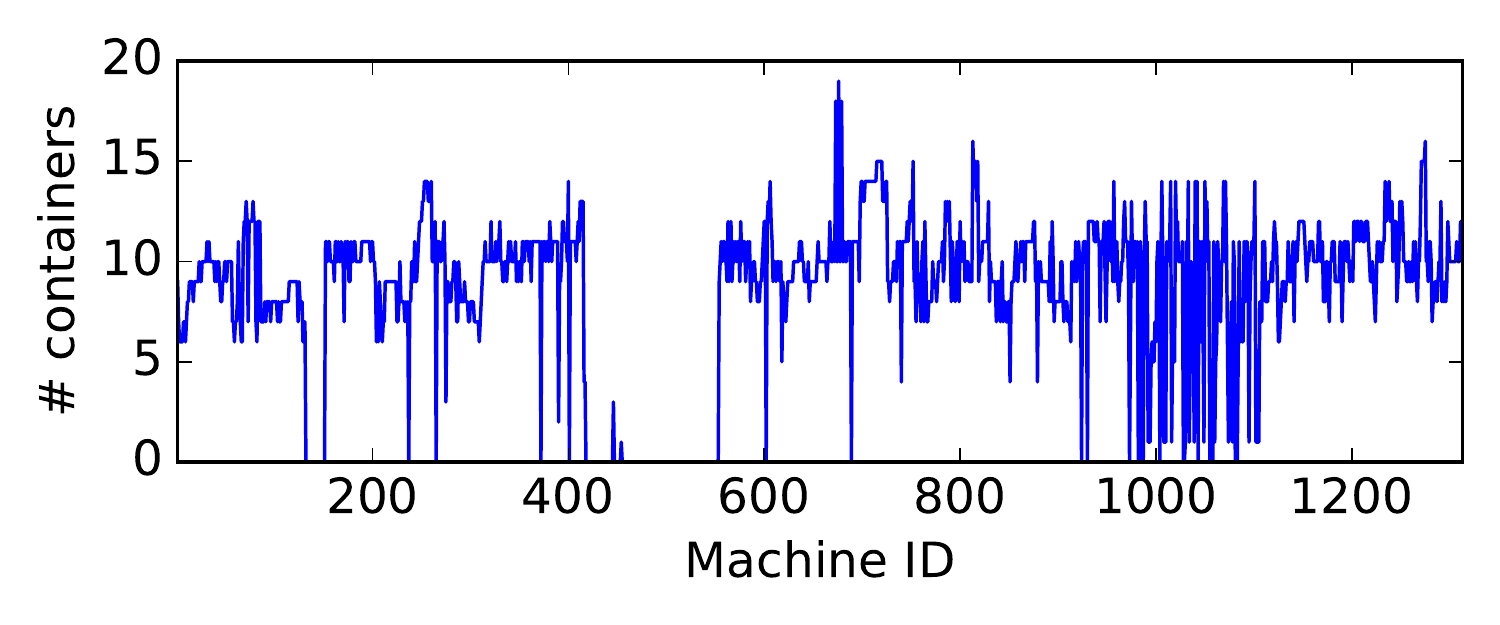}
\vspace{-15pt}
\caption{{\small
Distribution of containers across the cluster.
}}
\vspace{-15pt}
\label{fig:c_dist}
\end{center}
\end{figure}

\textbf{Resource reservation.}
We first calculate the per-machine resource amount requested and
reserved by containerized long-running jobs from the container
request trace ({\small\tt{container\_event.csv}}). 
Figure~\ref{fig:c_alloc} show that, 
except for a few servers with overbooked CPUs (spikes in machine
671--679, and semi-spikes in machine 797--829), 
the machine-level CPU allocation are consistently capped at 64--the
maximum number of CPU cores of one machine.
Figure~\ref{fig:lr_cpu_req_dist} and Figure~\ref{fig:lr_mem_req_dist}
plot the same trend: for both CPU and memory allocation, overbooking
is rare but does exist; about half of the machines (637) get all 64
cores reserved for containerized long-running applications;
$60$--$80\%$ of memory on $74\%$ machines are reserved for
containers.
Deep troughs in Figure~\ref{fig:c_alloc} are due to zero container
deployment on the machines.
In fact, container management systems such as
Sigma~\cite{alibaba_sigma} need to consider a lot of other
constraints when performing scheduling for long-running containers,
including affinity and anti-affinity constraints (e.g., co-locating
applications that belong to the same services for reducing network
cost, or co-locating applications with complementary runtime
behaviors), application priorities, whether or not the co-located
applications tolerate resource overbooking of the same host
machine~\cite{medea_eurosys18}.
A side effect of such multi-constraint multi-objective optimization
is that
the number of containers is unevenly distributed across the cluster
as shown in Figure~\ref{fig:c_dist}. 

\begin{figure}[t]
\begin{center}
\subfigure[CPU usage.] {
\includegraphics[width=.235\textwidth]{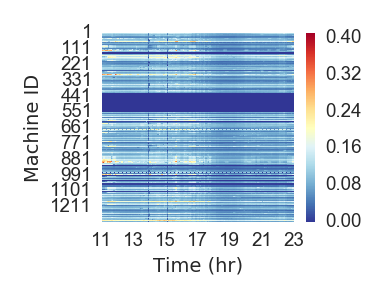}
\label{fig:c_cpu}
}
\hspace{-1.5em}
\subfigure[Memory usage.] {
\includegraphics[width=.235\textwidth]{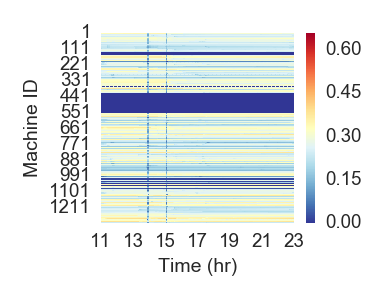}
\label{fig:c_mem}
}
\vspace{-15pt}
\caption[]{{\small
Container job resource usage (hour~11--23).
}}
\label{fig:c}
\end{center}
\vspace{-15pt}
\end{figure}

\textbf{Resource usage.}
The container workload trace ({\small\tt{container\_usage.csv}}) samples the
resource usage of each container every 5 minutes. We aggregate all
the container-level resource usage statistics by the machine ID based
on $container\rightarrow machine\_ID$ mapping recorded in the
{\small\tt{container\_event.csv}} file and plot the resource usage
heat map shown in Figure~\ref{fig:c}. The dominating pattern
is the horizontal stripes across the 12-hour tracing period\footnote{
The wide dark blue stripe showing no container activity is due to
that Alibaba intentionally leave a portion of machines as buffer area
which, if the long-running applications (containers) are of low load,
are solely designated for running batch jobs. When online service are
experiencing high load, batch jobs running in buffer area can be
evicted and resources can be preempted by the
containers~\cite{alibaba_sigma}.
}. Each stripe corresponds to one machine holding
multiple containers, clearly reflecting the long-running
nature\footnote{ Container re-scheduling and migration is not fully
supported in Sigma yet.  
} of containerized applications.
Another observation we make is that, even though Sigma tries to
balance out the amount of reserved resources (as one constraint of
the scheduling heuristic), the actual CPU and memory usage by
container workloads are imbalanced across all machines.

\begin{figure}[t]
\begin{center}
\subfigure[CPU.] {
\includegraphics[width=.45\textwidth]{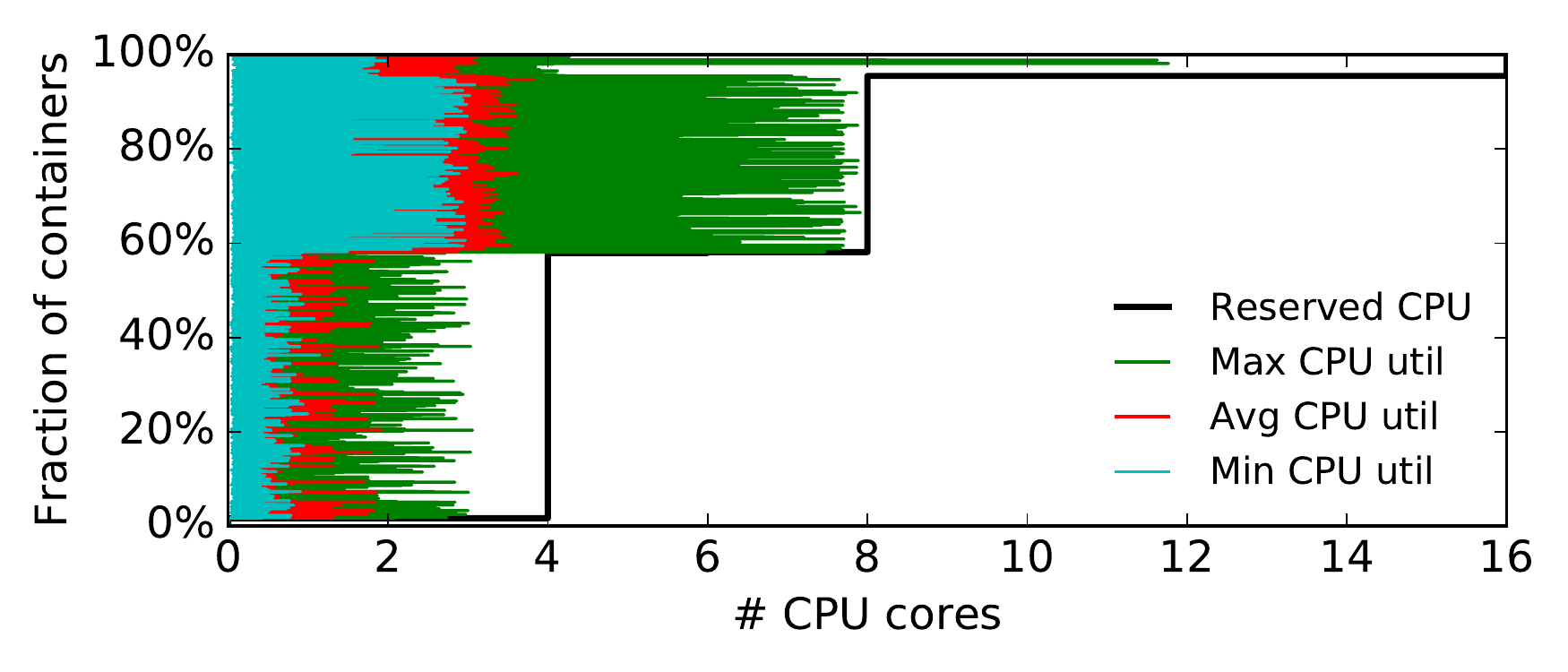}
\label{fig:c_cpu_dyna}
}
\subfigure[Memory.] {
\includegraphics[width=.45\textwidth]{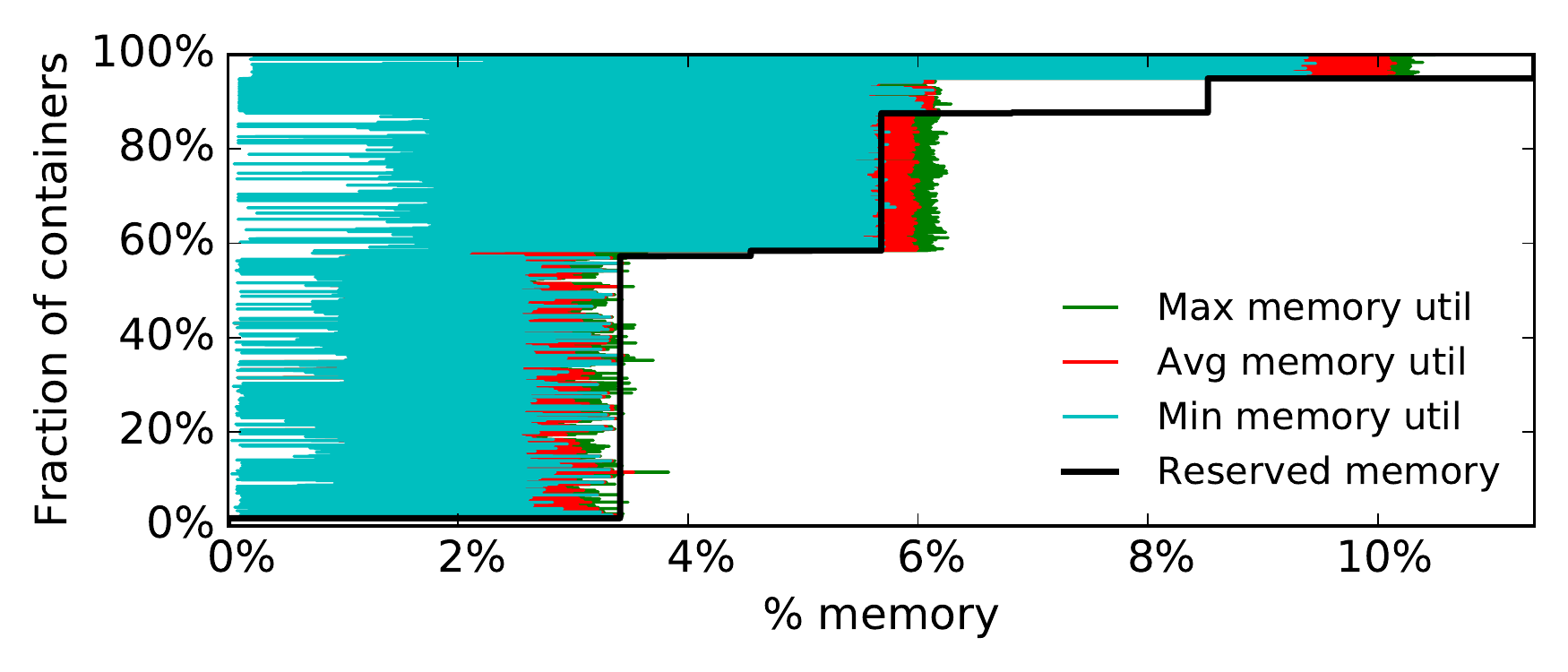}
\label{fig:c_mem_dyna}
}
\vspace{-15pt}
\caption{{\small
Reserved CPU and memory vs. their actual usage. 
We break down each container's
12-hour usage changes into \{Min, Avg, Max\} to show the dynamicity. 
The X-axis is sorted by the reserved resource amount.
}}
\label{fig:c_dyna}
\end{center}
\vspace{-15pt}
\end{figure}

\textbf{Resource Over-provisioning and Usage Dynamicity.}
To better understand the observed resource usage imbalance, we
compare the reserved CPU and memory capacity with the actual usage,
as shown in Figure~\ref{fig:c_dyna}.
We make the following observations:
(1)~CPU resources are over-provisioned by all containers
(Figure~\ref{fig:c_cpu_dyna}), while memory resources are
over-committed by a large majority of containers
(Figure~\ref{fig:c_mem_dyna}).
(2)~The CPU and memory request patterns are clearly visible---CPU
requests have 4 distinguishable patterns while memory requests have 6.
(3)~$\sim60\%$ of containers are inactive in terms of CPU usage,
having less than $1\%$ average CPU utilization with the maximum
percentile capped at $3\%$; average resource usage correlates to
temporal stability---the more resource consumed on average, the
higher the temporal variation tends to be---this is especially true
for CPU; memory usage dynamicity is not as high as that of CPU.
(4)~Most containers have a higher average memory usage above $2\%$,
which is as expected since containers need a minimum amount of memory
to keep online services functional.
This is, in fact, consistent with the well-studied behaviors of
web-scale distributed storage workloads~\cite{docker_fast18,
mbal_eurosys15, memshare_atc17, mos_pdsw15, mos_hpdc16}.

\textbf{Insights.}
Based on the observations, we infer the following.
(1)~A majority of the long-running containerized interactive services
stay inactive (at least during the 12-hour trace period); this
finding is consistent with other interactive workload
studies~\cite{memcached_sigmetrics12,memcache_nsdi13}\footnote{
E.g., Facebook's largest Memcached pool {\small\texttt{ETC}} deploys
hundreds of Memcached servers but only absorbs an incredibly low
average of 50K queries per second~\cite{memcached_sigmetrics12}.
}, implying a demand for elastic interactive service systems.
(2)~Long-running services are relatively more memory-hungry; ample
CPU resources reserved through resource over-provisioning are needed
to provide performance guarantee (stringent latency and throughput
requirement); this is especially true for in-memory computing whose
first-priority resources are essentially memory\footnote{
Being consistent with the observations made in the Memcached workload
study~\cite{memcached_sigmetrics12}, companies like Facebook use a
large Memcached pool driven by the huge demand for large memory
capacity.
}.
(3)~It is possible to make accurate resource usage prediction based
on the temporal usage dynamicity profiling~\cite{rc_sosp17},
especially for memory which has a relatively stable usage pattern
(see Figure~\ref{fig:c_dyna}(b)); the prediction can be used for more
informed resource management decision making such as container
re-scheduling/migration.

%% file: batch.tex
\pdfoutput=1

\vspace{-8pt}
\section{Batch Job Workloads}
\label{sec:batch}

\begin{figure}[t]
\begin{center}
\subfigure[CPU usage.] {
\includegraphics[width=.235\textwidth]{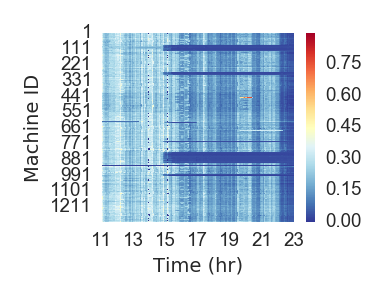}
\label{fig:b_cpu}
}
\hspace{-1.5em}
\subfigure[Memory usage.] {
\includegraphics[width=.235\textwidth]{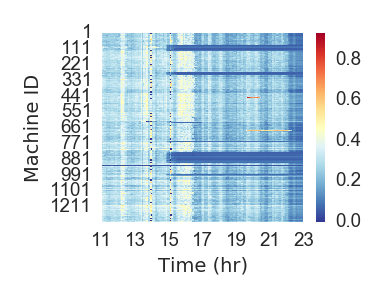}
\label{fig:b_mem}
}
\vspace{-15pt}
\caption{{\small
Batch job resource usage (hour~11--23).
}}
\label{fig:b}
\end{center}
\vspace{-15pt}
\end{figure}

\textbf{Resource usage.}
The batch job workloads exhibit different resource patterns compared
to that of containerized long-running job workloads. We calculate the
batch job workload resource usage shown in Figure~\ref{fig:b} by
subtracting the usage of containers (Figure~\ref{fig:c}) from the
overall usage (Figure~\ref{fig:overall}) of the cluster. We confirmed
the accuracy of the results by comparing it against the sum of all
batch task instances' runtime usage values under a certain timestamp.
The vertical stripes
(batch job waves) in Figure~\ref{fig:b} is due to the dynamic nature
of batch job workloads---task instances are transient and most of
them finish in seconds.  We can also observe that, within a single
wave, the CPU and memory resource usage are roughly balanced across
the cluster (later quantitatively demonstrated in Section~\ref{sec:col}),
except for some regions with no batch jobs scheduled (i.e., the
horizontal, dark blue stripes). 
This is because:
(1)~Fuxi is not constrained by data locality thanks to compute
and storage disaggregated infrastructure at Alibaba (for batch jobs all
data are stored and accessed remotely~\cite{alistore}), hence task
instances can be scheduled anywhere there is enough resource\footnote{
Batch job trace has no disk statistics capturing intermediate data
storage usage.
};
(2)~Fuxi adopts an incremental scheduling heuristic that
incrementally fulfills the resource demands at per-machine
level~\cite{fuxi_vldb14}.


\begin{figure}[t]
\begin{center}
\subfigure[Avg vs. requested CPU.] {
\includegraphics[width=.21\textwidth]{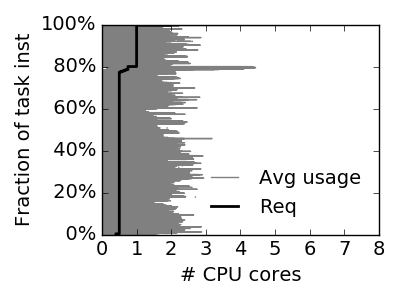}
\label{fig:b_cpu_cdf}
}
\hspace{-1em}
\subfigure[Avg/requested CPU ratio.] {
\includegraphics[width=.21\textwidth]{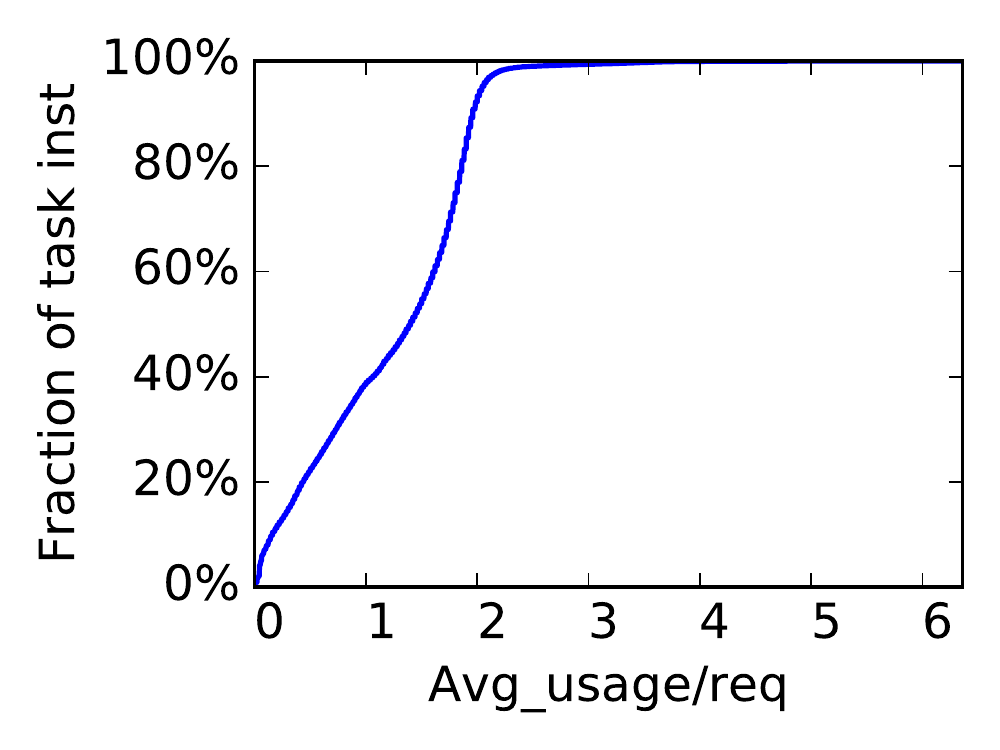}
\label{fig:b_cpu_ratio}
}
\subfigure[Avg vs. requested memory.] {
\includegraphics[width=.21\textwidth]{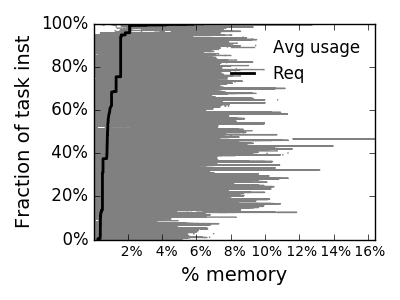}
\label{fig:b_mem_cdf}
}
\hspace{-1em}
\subfigure[Avg/requested memory.] {
\includegraphics[width=.21\textwidth]{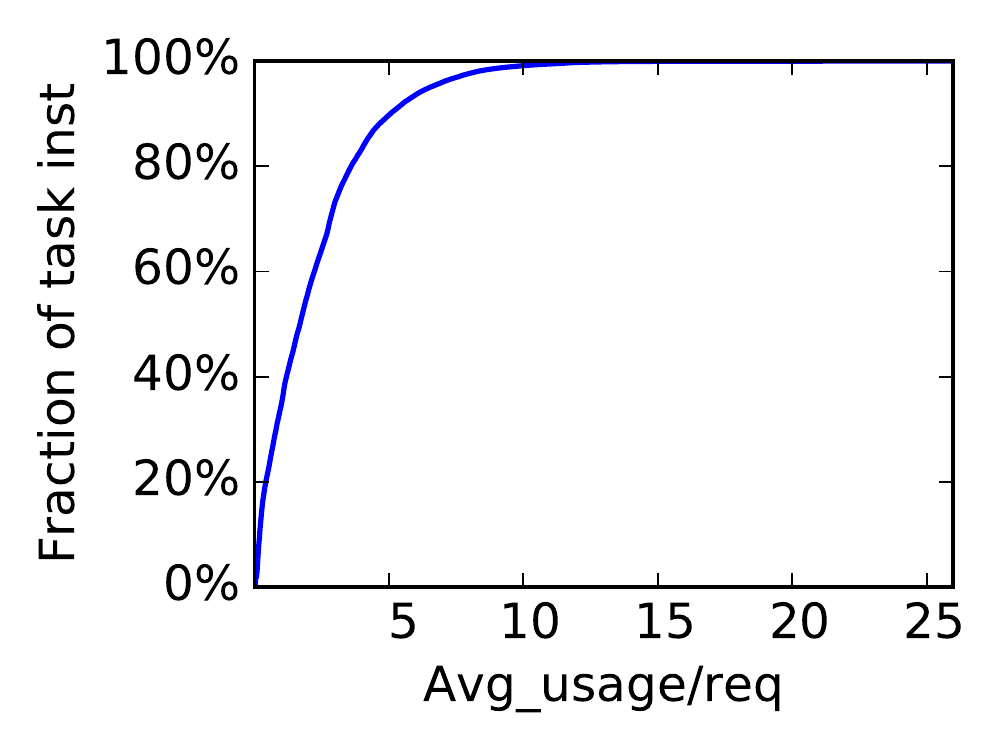}
\label{fig:b_mem_ratio}
}
\vspace{-15pt}
\caption{{\small
CDF of the average CPU and memory usage relative to the resource
request for the corresponding task instance.
}}
\label{fig:b_cdf}
\end{center}
\vspace{-15pt}
\end{figure}

Figure~\ref{fig:b_cdf} depicts the CDFs of requested CPU amount vs.
runtime CPU usage.
The over-commitment trend is clearly shown in
Figure~\ref{fig:b_cpu_cdf} and Figure~\ref{fig:b_mem_cdf}.
Quantitatively, $\sim30\%$ task instances use CPU cores more than
requested (Figure~\ref{fig:b_cpu_ratio}), while more than $70\%$ task
instances have an overcommitment ratio ($avg\_usage:req$ ratio) from
1--5 (Figure~\ref{fig:b_mem_ratio}).
This is understandable, as 
a large number of short-task-dominated, transient batch jobs can
\emph{elastically over-commit} resources that are originally reserved
by the \emph{over-provisioned long-running jobs}.

\textbf{Task scheduling.}
The trace files {\small\tt{batch\_task.csv}} and
{\small\tt{batch\_instance.csv}} contain the detailed batch job
profile information including job composition (how many tasks per job
and how many instances per task), spacial (how task instances are
mapped to the machines) and temporal information (how long each task
instance runs), and average/max resource usage (though not complete).
By combining the spacial/temporal and job composition information, we
can easily infer the DAG structure of a specific batch job. 

\begin{figure}[t]
\begin{center}
\subfigure[Execution waves.] {
\includegraphics[width=.23\textwidth]{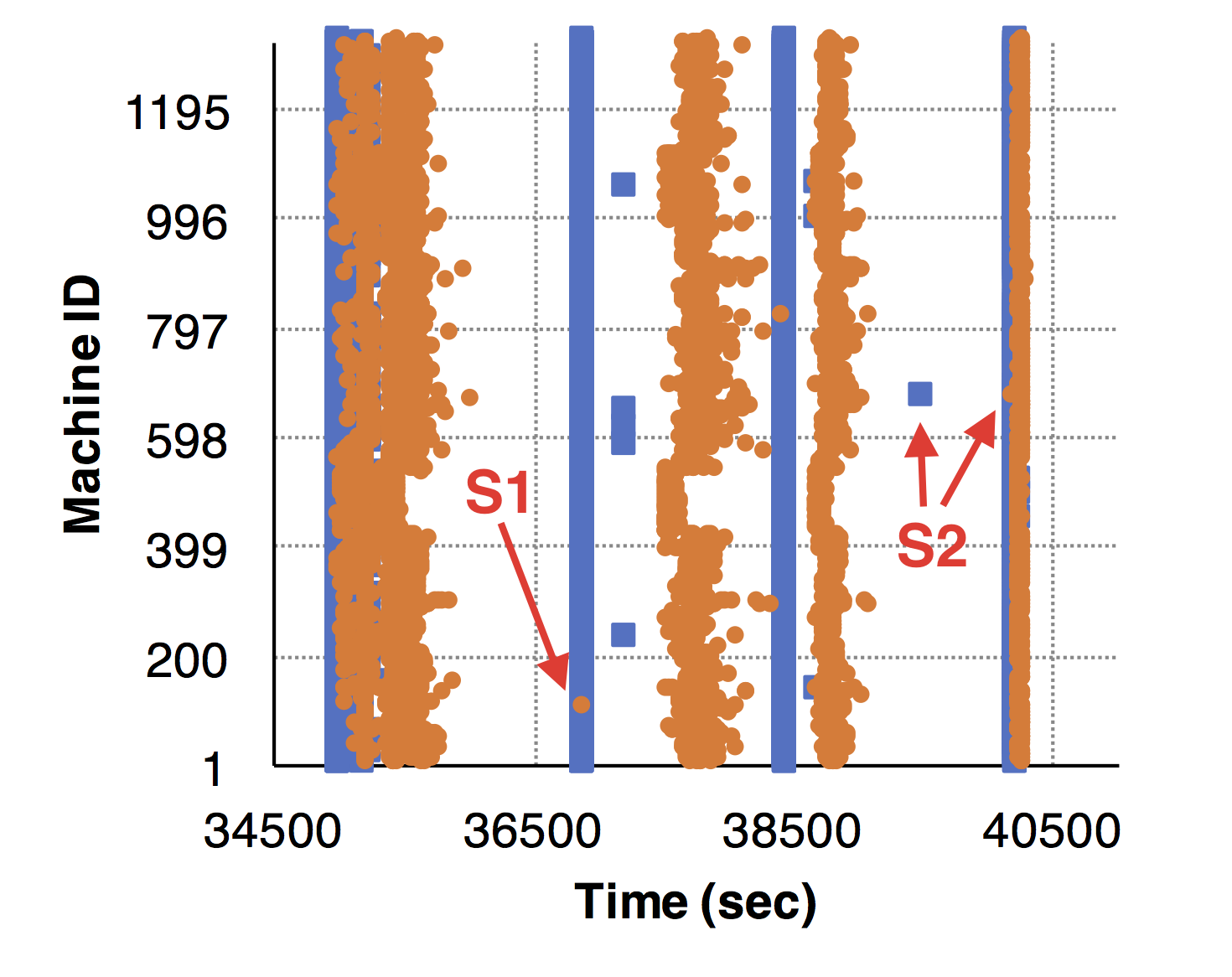}
\label{fig:wave}
}
\hspace{-1.8em}
\subfigure[DAG.] {
\includegraphics[width=.23\textwidth]{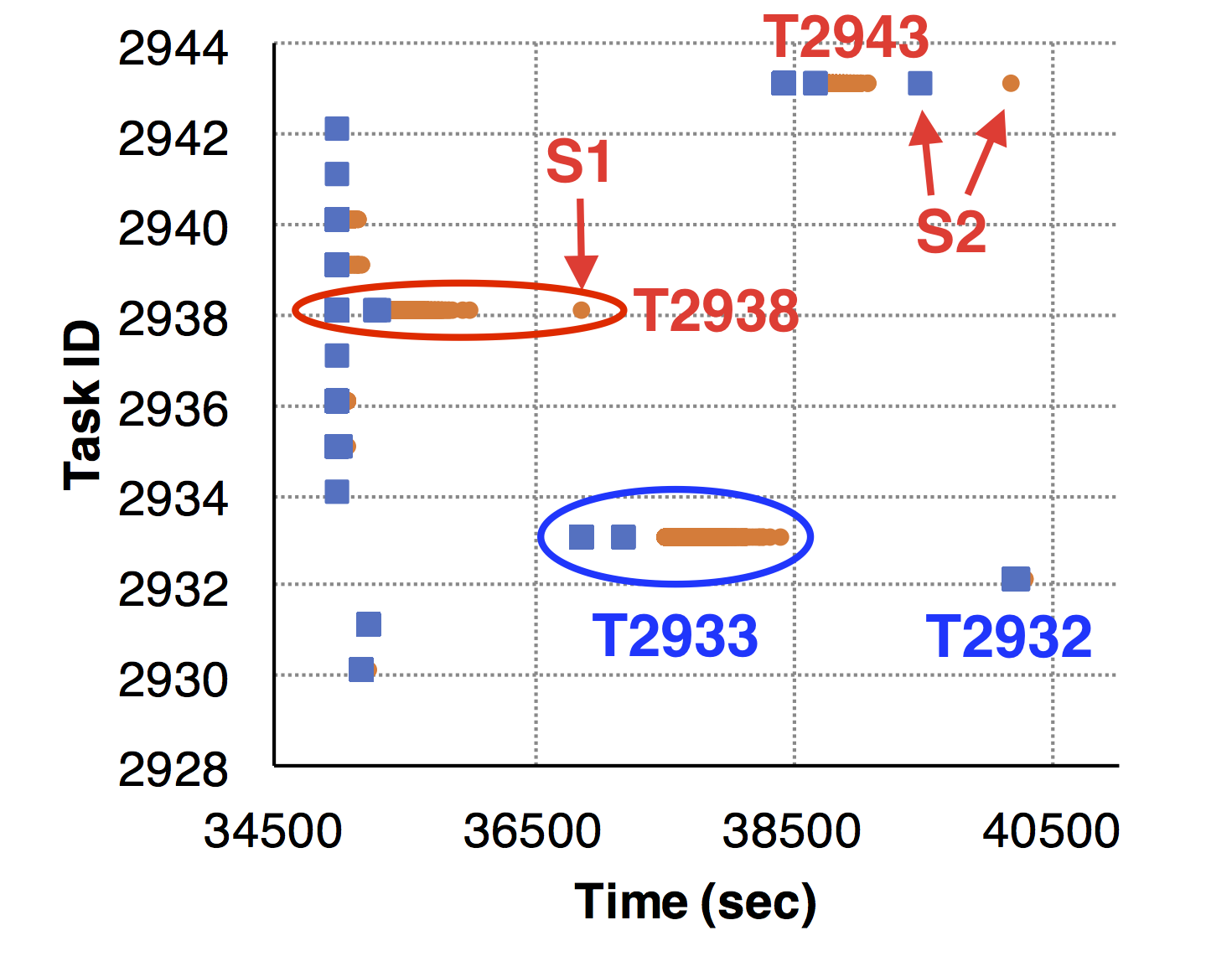}
\label{fig:dag}
}
\vspace{-15pt}
\caption{{\small
Job~556's execution profile; blue squares correspond to the start of
the makespan for a specific task instance, and orange circles
correspond to the end of the makespan.  
Figure~\ref{fig:wave} depicts the task waves with Y-axis showing the
machine IDs. Figure~\ref{fig:dag} plots the job DAG with Y-axis
showing the task IDs. 
{\small\tt{S1}}: Straggler~1; 
{\small\tt{S2}}: Straggler~2;
{\small\tt{T2933}}: Task~2933.
We can infer the DAG dependencies as follows: T2932 waits for T2943
to complete; T2943 depends on T2933; T2933 waits for all tasks of the
$1^{st}$ wave to finish.
}}
\label{fig:j556}
\end{center}
\vspace{-15pt}
\end{figure}

One question we want to answer is whether the well-studied
issue~\cite{dolly_nsdi13, late_osdi08} still persists in Alibaba's
datacenters--The answer is surprisingly yes. 
To illustrate the impact of stragglers on job performance,
Figure~\ref{fig:j556} visualizes Job~556's execution.
We can easily spot two stragglers. {\small\tt{S1}}, which has a
starting timestamp same as the rest other instances of the same task
{\small\tt{T2938}} but an end timestamp way behind the rest, results
in delayed scheduling and execution of {\small\tt{T2933}} (the
$2^{nd}$ wave in Figure~\ref{fig:wave}). {\small\tt{S2}}, belonging
to {\small\tt{T2943}}, has a lagged starting timestamp greater than
that of the rest; a direct result is the delayed execution of
{\small\tt{T2932}} (the $4^{th}$ wave in
Figure~\ref{fig:wave}). To distinguish between these two typical
cases, we call the {\small\tt{S1}} kind of task instances as
\textbf{stragglers}, and the {\small\tt{S2}} kind of tasks instances
\textbf{starvers}.

\begin{figure}[t]
\begin{center}
\subfigure[Straggler ratio distribution.] {
\includegraphics[width=.23\textwidth]{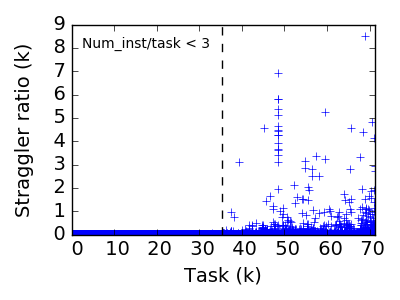}
\label{fig:straggler}
}
\hspace{-1.2em}
\subfigure[CDF of straggler ratios.] {
\includegraphics[width=.23\textwidth]{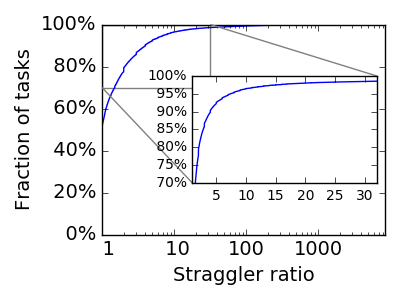}
\label{fig:straggler_cdf}
}
\subfigure[Starvation delay distribution.] {
\includegraphics[width=.23\textwidth]{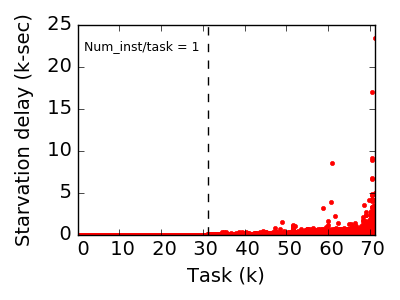}
\label{fig:starvation}
}
\hspace{-1.2em}
\subfigure[CDF of starvation delays.] {
\includegraphics[width=.23\textwidth]{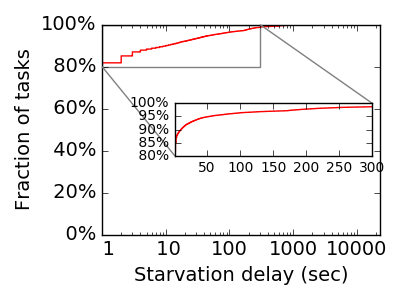}
\label{fig:starvation_cdf}
}
\vspace{-15pt}
\caption{{\small
Straggler and starver analysis. In Figure~\ref{fig:straggler} and
Figure~\ref{fig:starvation}, the X-axis is sorted smallest to largest
by the number of instances per task. 
}}
\label{fig:perf}
\end{center}
\vspace{-15pt}
\end{figure}

We then scan the execution profiles of all tasks included in the
trace. We iterate through all tasks and calculate the \emph{straggler
ratios} (defined as the ratio of the maximum and minimum instance
makespan of the corresponding tasks) and \emph{starvation delays}
(defined as the difference between the largest and the smallest
instance starting timestamp of the corresponding tasks).
Figure~\ref{fig:straggler} and Figure~\ref{fig:starvation} depict the
straggler and starver distribution, respectively.
As the task size (i.e., number of instances per task) increases, the
straggler ratio and starvation delay increases accordingly. 
As shown in Figure~\ref{fig:straggler_cdf}, around $50\%$ tasks have
a straggler ratio of 1, because half of them have 1 or 2 instances.
$\sim7\%$ tasks have a straggler ratio greater than $5\times$, with
the highest as $8522\times$.
Figure~\ref{fig:starvation_cdf} shows that $\sim4\%$ tasks have a
starvation delay longer than 100~seconds, with the longest as
23379~seconds. 

We further investigate the reasons 
by looking into the detailed task traces. Straggler patterns included
in the trace and possible causes are classified as follows.
(1)~Some straggler is significantly slower than the rest, which seems
a common case due to either misconfiguration (e.g., disabling
speculative execution~\cite{spec}) or severely imbalanced load (e.g., a
task instance getting more work to do than the rest).
(2)~Some task instances have long execution time and hence suffer a
higher chance of getting failed; failed instances get re-scheduled
(caught in the trace) and as a result suffer long starvation delay.
(3)~A straggler 
is spotted and interrupted\footnote{
Fuxi's speculative execution technique where Fuxi interrupts
long-tail instance and launches backup instance~\cite{fuxi_vldb14}.
} at a very late time (one extreme example in the trace: few hours
after $99\%$ of task instances of have finished the execution, while
most of other task instances finish in seconds)
by Fuxi; as a result, Fuxi launches a speculative backup instance,
which may finish quick or take very long time to finish; either way,
this scenario results in a false positive long starvation delay, or
much worse--an extremely large straggler ratio.
(4)~Some task instance is interrupted and gets indefinitely starved
(possibly due to low priority) while waiting for re-scheduling,
resulting in surprisingly long starvation delay.

\textbf{Insights.}
Based on the observations, we infer the following.
(1) Workloads of transient batch jobs with many short-lived tasks
exhibit resource patterns perfectly complementing that of
over-provisioned, long-running, mostly CPU-inactive, containerized
applications; those free resources not yet consumed by the long-running
applications have to be efficiently utilized by batch jobs;
essentially, this is key to improve the overall cluster utilization;
furthermore, users do not quite care about accurately estimating the
actual resource usage of the to-be-submitted batch jobs\footnote{
Our hypothesis got confirmed by Alibaba engineers.
};
(2)~The notoriously persisting straggler issues might be eliminated by
enforcing/adding more comprehensive configuration settings at the
administrator/developer side or via more careful data partitioning
planning at the user side.  For example, 
for interrupted task instances that are suffering from starvation
(Fuxi failed to re-schedule them in time), some priority-based
scheme (e.g., MLFQ-liked scheduling heuristic~\cite{mlfq}) may help.
Users also take responsibilities.  Users should carefully plan on
data partitioning to avoid imbalanced input assignments.

%% file: colocation.tex
\pdfoutput=1

\section{Workload Co-location}
\label{sec:col}

\begin{figure}[t]
\begin{center}
\subfigure[CPU usage.] {
\includegraphics[width=.235\textwidth]{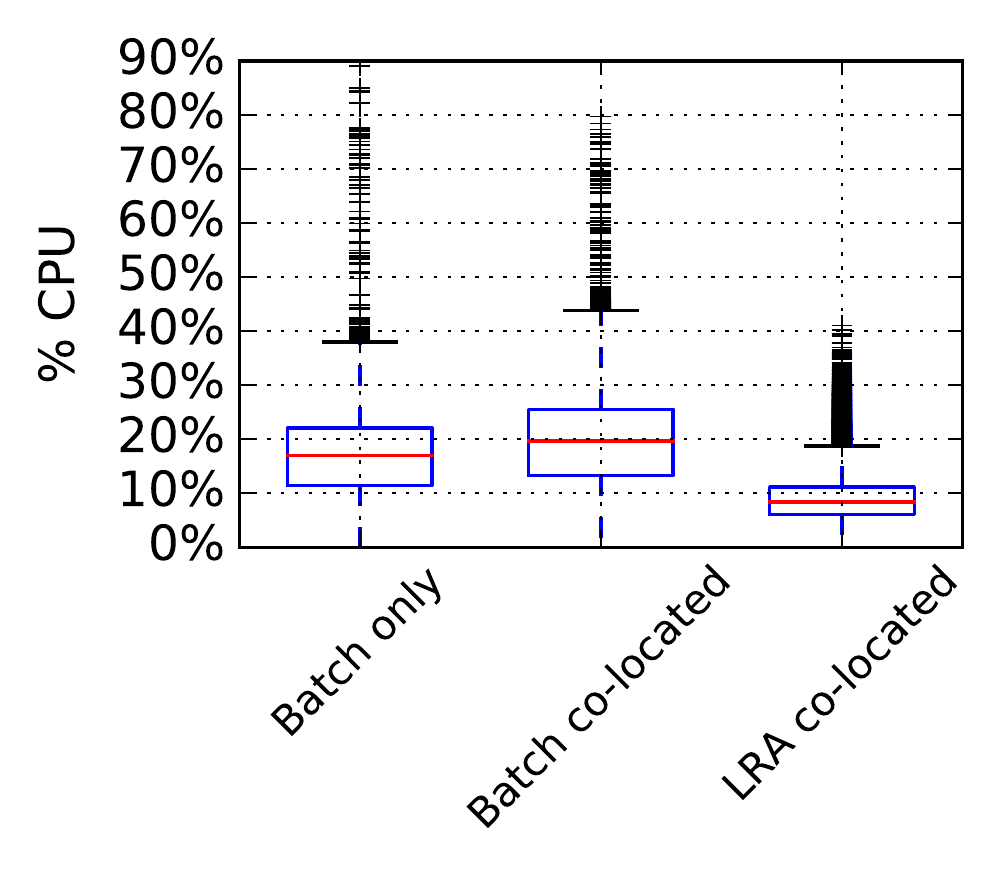}
\label{fig:col_cpu}
}
\hspace{-1.4em}
\subfigure[Memory usage.] {
\includegraphics[width=.235\textwidth]{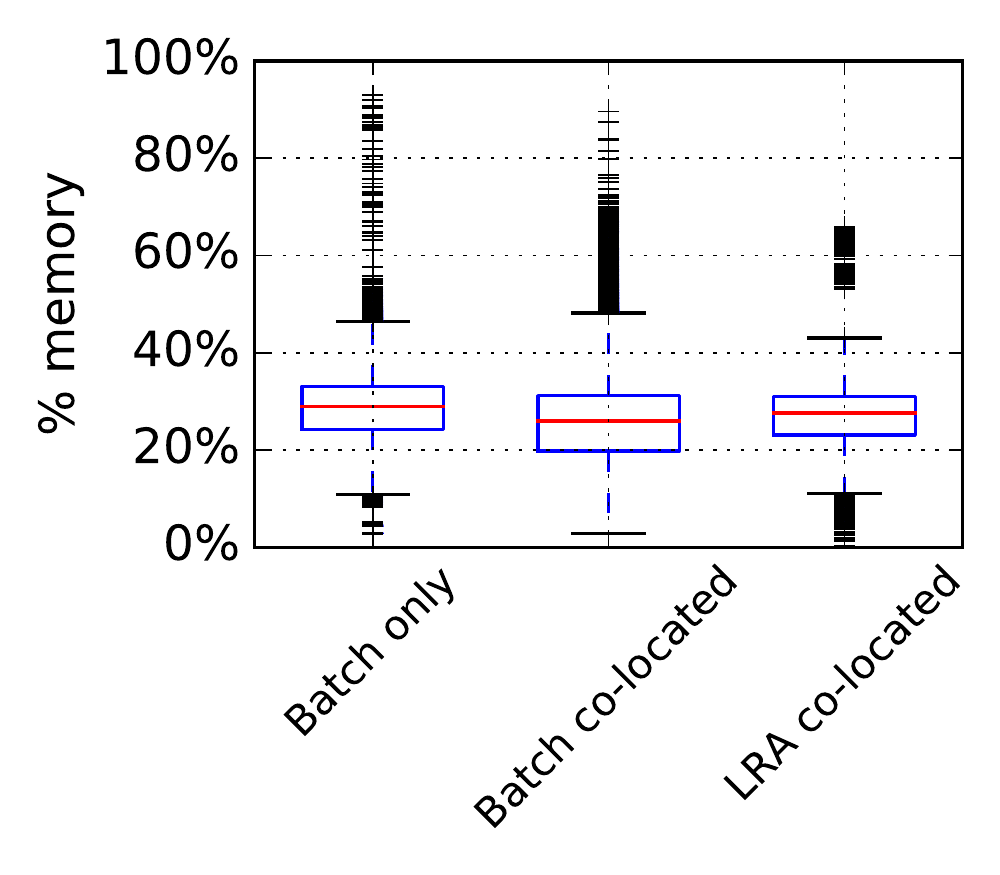}
\label{fig:col_mem}
}
\vspace{-5pt}
\caption{{\small
Box-and-whisker plots showing CPU and memory usage distributions.
{\small\texttt{Batch only}}: the machine region hosting batch jobs only;
{\small\texttt{Batch co-located}}: batch jobs' resource usage in region  
hosting both batch and long-running applications (LRA);
{\small\texttt{LRA co-located}}: LRA's resource usage in region hosting both.
}}
\label{fig:col_usage}
\end{center}
\vspace{-10pt}
\end{figure}

\textbf{Resource usage}
Recall that Figure~\ref{fig:c} shows machine region in the cluster
with no container deployment (the dark blue horizontal stripe). We
are particularly interested in how Fuxi~\cite{fuxi_vldb14} allocates
resources in such {\small\verb+Batch only+} machine region. We thus
partition the cluster into a {\small\verb+Batch only+} region where
only batch jobs are running, and a {\small\verb+Co-located+} region
where both long-running and batch jobs are sharing the resources.
Figure~\ref{fig:col_usage} depicts the CPU and memory resource usage
distribution as a function of workload type and partitioned machine
region.
We observe that in {\small\verb+Batch only+} region the average
resource utilization is almost the same as that of batch jobs in
{\small\verb+Co-located+} region.
This implies that: (1)~{\small\verb+Batch only+} region's resource
utilization is significantly lower than that of the co-located
region; and (2)~Fuxi--the batch scheduler--does not take into account
the resource usage heterogeneity caused by co-located long-running
job workloads.

\begin{figure}[t]
\begin{center}
\subfigure[CPI.] {
\includegraphics[width=.235\textwidth]{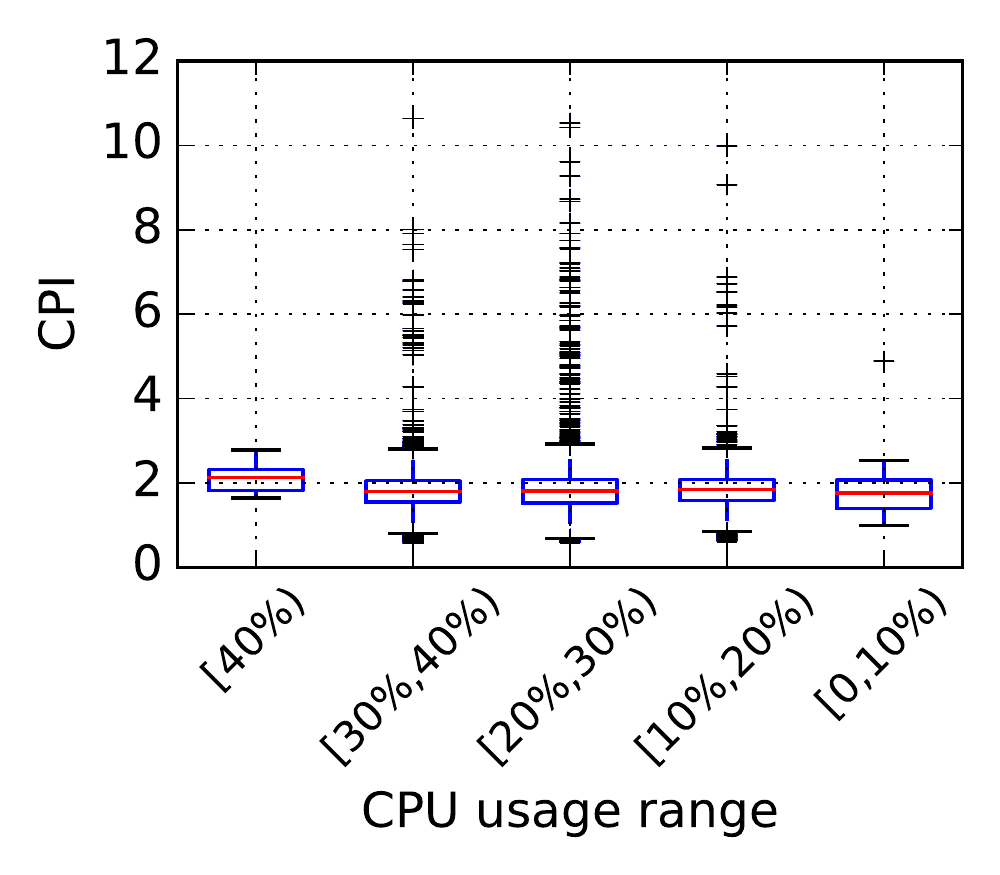}
\label{fig:col_cpi}
}
\hspace{-1.4em}
\subfigure[MPKI.] {
\includegraphics[width=.235\textwidth]{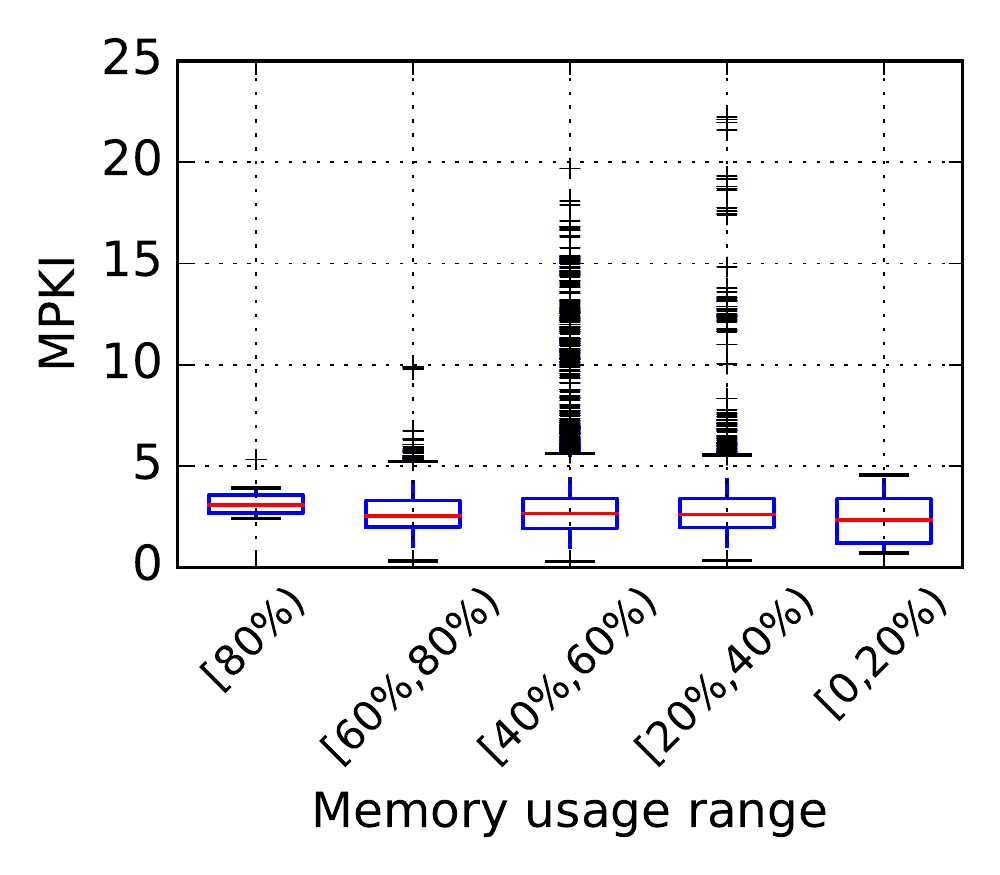}
\label{fig:col_mpki}
}
\vspace{-5pt}
\caption{{\small
Box-and-whisker plots showing maximum CPI and MPKI distribution as a
function of machine's resource usage range (for both CPI and MPKI the
lower the better).
}}
\label{fig:col_perf}
\end{center}
\vspace{-10pt}
\end{figure}

\textbf{Performance metrics}
The container trace records the runtime performance metrics including
mean/maximum CPI (cycles per instruction) and MPKI (memory
accesses per kilo-instructions). To study the container co-location
impact on performance, we break down the 
CPU and memory resource usage into ranges and partition the cluster
based on that. Figure~\ref{fig:col_perf} plots the maximum CPI and
MPKI distributions at different resource usage ranges at per machine
level. Statistically, both CPI and MPKI (the major percentiles e.g.
medium) reaches the highest at highest resource utilization: $40\%$+
for CPU usage, and $80\%$+ for memory usage.  Outliers at other usage
ranges do exist and account for only a negligible set of data points.
Note that the resource usage here accounts for both containerized
long-running applications and non-containerized batch job workloads.
This also implies that co-location tends to introduce performance
interference when resource contention (i.e., resource usage)
increases.

\textbf{Insights}
Based on the observations, we infer the following.
Fuxi makes seemingly independent scheduling decisions by assuming
a homogeneous resource pool, regardless of the co-existence of the
long-running container deployment; however, the heterogeneity caused
by Sigma should be hinted via the global Level-0
controller~\cite{alibaba_sigma} to Fuxi for more efficient resource
scheduling. For example, a smart global controller would be able to
detect low resource utilization and take action by accommodating more
batch jobs at {\small\verb+Batch only+} region.
\emph{Multiple workload-specific resource schedulers demand a better
global controller design that cohesively manages complex co-located
workloads}---this may be directly applicable to Alibaba's two-level
CMS architecture; we argue, however, that the generality of the
insights is critical for system designers and IT practitioners
working on CMSs as well.

%% file: conclusion.tex
\pdfoutput=1

\section{Conclusion, Discussion Points, and Future Work}

Aiming at improving the overall resource utilization, workload
co-location results in exponentially increased complexity for
warehouse-scale datacenter resource management.  Analysis of the
Alibaba cluster trace reveals such challenges, for which new resource
scheduling approach that has a deeper sense of co-location will
likely be necessary. We believe that our findings will lead to hot
discussion on the following points of interest.
(1)~How to improve current global coordinator design so as to
seamlessly incorporates the diversified but complementary workload
behaviors of co-located workloads for more efficient resource
management?
(2)~What machine learning (ML) algorithms are most suitable and how
much training data is needed for accurately predicting the resource
dynamicity and footprint of container workloads or even the highly
dynamic batch workloads with interferences~\cite{rc_sosp17}? 
(3)~Potential issues introduced to designers of CMSs when co-locating
more than 2 types of workloads 
with diversified dynamicity and resource usage patterns.  A
particular point of interest is co-locating and managing large-scale
ML workloads~\cite{tensorflow_osdi16, proteus_eurosys17, slaq_socc17}
with intensive use of heterogeneous
accelerators~\cite{geeps_eurosys16, tetrisched_eurosys16}.
In the future, we would like to investigate and explore the above
directions as well.

\paragraph*{Acknowledgments}
{
We thank Dr. Haiyang Ding from Alibaba for his valuable feedback on
an early version of this manuscript.
This work was sponsored by GMU.
}